\documentclass{article}
\usepackage{spconf,amsmath,graphicx}
\usepackage{multirow}
\usepackage{multicol}
\usepackage{cite,balance}
\title{ADAPTIVE-AVG-POOLING BASED ATTENTION VISION TRANSFORMER\\ FOR FACE ANTI-SPOOFING}
\name     {Jichen Yang$^1$, Fangfan Chen$^1$, Rohan Kumar Das$^2$, Zhengyu Zhu$^{1,*}$, Shunsi Zhang$^3$}
\address  {$^1$School of Cyber Security, Guangdong Polytechnic Normal University, Guangzhou, China\\
           $^2$Fortemedia Singapore, Singapore\\
           $^3$Guangzhou Quewan Network Technology Co. Limited Ltd., Guangzhou, China
\thanks{*Corresponing author. This work was supported in part by the Science and Technology Program (Key R\&D Program) of Guangzhou, China (2023B01J0004), special projects in key areas of Guangdong Provincial Department of Education (2023ZDZX1006) and Research project of Guangdong Polytechnic Normal University, China (2023SDKYA019).}}
           
\begin{document}
%\ninept
%
\maketitle
\begin{abstract} 
Traditional vision transformer consists of two parts: transformer encoder and multi-layer perception (MLP). The former plays the role of feature learning to obtain better representation, while the latter plays the role of classification. Here, the MLP is constituted of two fully connected (FC) layers, average value computing, FC layer and softmax layer. However, due to the use of average value computing module, some useful information may get lost, which we plan to preserve by the use of alternative framework. In this work, we propose a novel vision transformer referred to as adaptive-avg-pooling based attention vision transformer (AAViT) that uses modules of adaptive average pooling and attention to replace the module of average value computing. We explore the proposed AAViT for the studies on face anti-spoofing using Replay-Attack database. The experiments show that the AAViT outperforms vision transformer in face anti-spoofing by producing a reduced equal error rate. In addition, we found that the proposed AAViT can perform much better than some commonly used neural networks such as ResNet and some other known systems on the Replay-Attack corpus.  
\end{abstract}
\section{Introduction}
\label{sec:intro}

In the recent years, the development of deep learning technologies have led to several applications. Among such applications, the systems developed for person authentication using biometrics have gained significant attention. Face, voice, iris and fingerprint are some common biometric measures that are used in real-world scenarios~\cite{FaceRecogICASSP,ijst_deb,IrisRecog}. There are also systems that use multi-modal biometric measures for development of robust systems. However, the biometric systems are vulnerable to various kinds of spoofing attacks~\cite{Biometric_spoof}. Especially, with advent of generative models, it has become much easier to generate spoofed data to attack any biometric systems.  

% In recent years, with deep-learning development, it becomes more and more easier to generate multimedia content. Furthermore, plenty of generated multimedia content such as spoofed speech and spoofed faces are generated every day.

% In order to prevent spoofed speech attack speaker verification system and boost the study of speech anti-spoofing, in 2015, 2017, 2019 and 2021,
% automatic speaker verification spoofing and countermeasure challenge (ASVspoof) has been held~\cite{asvspoof2015, asvspoof2017, asvspoof2019, asvspoof2021}, respectively.

% According to spoofed speech type, spoofed speech detection can be divided into synthetic speech detection~\cite{asvspoof2015} and replay speech detection~~\cite{asvspoof2017}. In which, synthetic speech consists of synthesized speech~\cite{voicecloning} and voice converted speech~\cite{ttsvc}. It is noted that both synthesized speech and voice converted speech are generated by computer while replay speech is generated by recording device and replay device, in other words, replay speech is usually obtained by applying recording device on genuine speech and replay it.

Face and voice are the most often used biometrics that encounter threat from spoofing attacks. In order to prevent such attacks, anti-spoofing systems for face or voice have attracted much research in the past decade~\cite{FAC_review,spoof_review,bib:Attacker_overview2020}. While voice data of speakers have more variability across sessions, face has comparatively less variability and thus the latter has an edge over the former modality. However, the face recognition systems are vulnerable to different presentation attacks such as replay, print, 3D-mask and makeup~\cite{FAC_review}. In this work, we focus on face anti-spoofing to prevent spoofing attacks performed on face recognition systems.

The earliest works for face anti-spoofing focused more on designing handcrafted features~\cite{Eyeblink_face, face_context,face_gen}. Such representations focus more on human liveness curves, eye blinking and face as well head movements. Later, advanced representations like deep learning based methods like end-to-end systems~\cite{e2e1,e2e2,e2e3} or hybrid methods that include handcrafted features with deep learning systems became the benchmark systems~\cite{hybrid1,hybrid2,hybrid3}. In the recent years, face anti-spoofing have attracted more attention for research and development of robust systems for applications~\cite{tmm2022,tpami2023,tifs2023dommain,icassp2023general,icassp2023partial,icassp2023liveness}. Among these, the works in~\cite{tifs2023dommain,icassp2023general,icassp2023partial} focus on improving the generalization capability of face anti-spoofing models. The authors of~\cite{tifs2023dommain} use negative data augmentation, while the authors of~\cite{icassp2023general} and ~\cite{icassp2023partial} use the learned causal representations and a partial domain-aware adaptation module, respectively.

With the advancements in deep learning methodologies, transformer was proposed for natural language processing~\cite{transformer} that soon became the state-of-the-art for various domains. The transformer architecture basically consists of two parts that are transformer encoder and transformer decoder. The transformer encoder has the function of feature learning, while transformer decoder can be used for the many tasks such as classification, regression by generating output sequence. Both transformer encoder and decoder have the same architecture, applying self-attention mechanism to learn useful features. As a result it was applied to different applications, a modified transformer referred to as vision transformer (ViT) was proposed for computer vision tasks~\cite{vit}. It is constituted of transformer encoder and multi-layer perception (MLP), where the former is used for feature learning to obtain excellent feature representation, while the latter is used for classification. It was found that ViT can achieve improved performance compared to the state-of-the-art convolutional neural networks as reported in~\cite{vit}.

In the traditional ViT, MLP contains a module called as average value computing between the two fully connected (FC) layers. Due to the use of average value computing module some information gets lost, which we believe to be useful for classification tasks, especially for presentation attacks, since non-target classes are designed by the attacker to have very close similarity to the target classes. In this work, we propose a modified ViT to overcome the loss of information by replacing the average value computing module by the use of adaptive-avg-pooling and attention. We refer this new transformer model as adaptive-avg-pooling based attention vision transformer (AAViT). The proposed AAViT is studied for face anti-spoofing using the Replay-Attack database.

\begin{figure} [t]
    \centering
    \includegraphics[width=8.7cm,height=12cm]{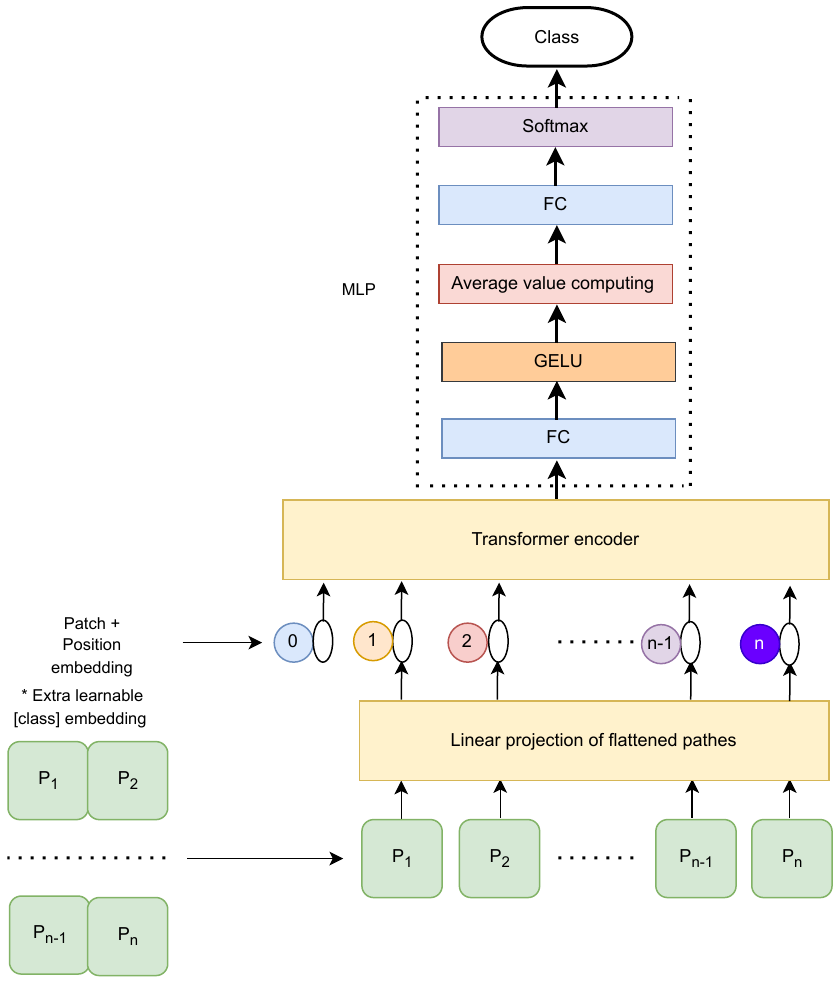}
    \caption{The diagram block of the traditional ViT.}
    \label{fig:vit}
\end{figure}

\section{Proposed AAViT for Face Anti-spoofing}
\label{secii}

In the section, we first introduce the traditional ViT~\cite{vit} and then on the basis of ViT, we describe the modified ViT, i.e., AAViT in detail.
\subsection{ViT}

%\begin{figure} [t]
%    \centering
%    \includegraphics[width=9cm,height=13cm]{fig/vitaa.pdf}
%    \caption{The diagram block of the proposed AAViT.}
%    \label{fig:vitap}
%\end{figure}

\begin{figure} [t]
    \centering
    \includegraphics[width=3.9cm,height=7.2cm]{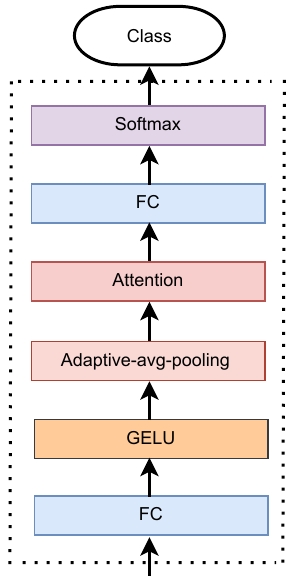}
    \caption{Block diagram of the proposed AAMLP module in AAViT.}
    \label{fig:aamlp}
\end{figure}

Fig.~\ref{fig:vit} shows the block diagram of the traditional ViT. As observed the original input image is first reshaped into $n$ patches and then linear projection is applied on the patches. In addition, the ViT has two parts: transformer encoder and MLP. The former is used for feature learning, while the latter is used for classification as discussed in the introduction.

The modules in MLP play their respective roles for classification task. The first FC is used to transform the input, then Gaussian error linear unit (GELU) module is used as the activation function, followed by average value computing is used to compute the average value of different dimensions. Then, the second FC is used to transform the number of one-dimension signals into the class number signals, and finally softmax is used to obtain the probability. 

% More details can be found in [13].

\subsection{AAViT}
%The diagram block of the proposed AAViT is given in Fig.~\ref{fig:vitap}. From Fig.~\ref{fig:vitap}, it can be found that the proposed AAViT mainly contains two parts: Transformer encoder and AAViT, further, there are six submodules in AAViT, which are two FCs, one GELU, one adaptive-avg-pooling, one attention and one softmax. The reason why we call the modified MLP as AAMLP is that the submodules of adaptive-avg-pooling and attention are used to modify tradition MLP. The same reason is that why we name the modified ViT as adaptive-avg-pooling and attention (AAViT).

%By comparing Fig.~\ref{fig:vit} and Fig.~\ref{fig:vitap}, it can be found that both AAViT and ViT have the Transformer encoder part, the difference between them is that MLP is used in ViT while  adaptive-avg-pooling and attention MLP (AAMLP) is used in AAViT. In other words, the module of average value computing in ViT is replaced by the modules of adaptive-avg-pooling and attention, then AAMLP is obtained.

%The module of Transformer encoder and the submodules of FC, GELU and softmax in AAViT play the same roles in those in ViT mentioned above. In addition, the module of adaptive-avg-pooling can preserve more texture information by reducing the shifting of average value caused by convolutional layer parameter error and then attention is used to give different weights for different dimensions of adaptive-avg-pooling results.

We now discuss about the proposed AAViT. It consists of two parts: the upper part is the modified MLP unlike that in original ViT. On the other hand, the lower part is transformer encoder, which is the same as that in Fig.~\ref{fig:vit}. We therefore have a close look at only the upper part of the proposed AAViT, which is shown in Fig.~\ref{fig:aamlp}. As observed from Fig.~\ref{fig:aamlp}, the module of average value computing in ViT is replaced by the modules of adaptive-avg-pooling and attention in AAViT. Further, as adaptive-avg-pooling based attention is used to modify the MLP part, we refer to it as the modified MLP or AAMLP in short. This naming convention is similar to the way we refer AAViT, which is modified ViT. 

% The proposed AAViT contains two parts: the upper part is modified MLP, the lower part is Transformer encoder, which us the same as the lower part in Fig.~\ref{fig:vit}. 

% The diagram block of the modified MLP is given in Fig.~\ref{fig:aamlp}. From Fig.~\ref{fig:aamlp}, it can be found that the module of average value computing is replaced by the modules of adaptive-avg-pooling and attention. Furthermore, because adaptive-avg-pooling based attention is used to modify MLP, we name the modified MLP is AAMLP for short. In the same way, we call the modified ViT as AAViT.

\section{Experiments}
\label{seciii}

In this section, we discuss the database, evaluation metric and the experimental setup in detail in the following subsections. 

\subsection{Database}

% To date, there are several types of FAS databases, which are print, replay, makeup and 3D-mask, wherein print and replay attacks are recorded in controlled indoor scenario while 3D-masks attacks are recorded in the complex conditions.

In this work, REPLAY-ATTACK\footnote{https://www.idiap.ch/en/dataset/replayattack} database is used for the studies. The database was produced by the IDIAP Institute in Switzerland and has three types of possible attacks using three different media and two different recording conditions~\cite{replayattack}. The three types of spoofing attacks to generate the spoofed faces are print-attack, phone-attack, and table-attack. The dataset is divided into training, development and test subsets, and customers (face of persons) in each subset do not appear in the others. The training set is used to train the anti-spoofing model, the development set is used to tune the model's parameters, the test set is used to report the performance. Table~\ref{tab:database} shows a summary of the subsets of REPLAY-ATTACK database containing real-access and three types of spoofing attacks.

% the number of videos of training, development and test subsets of REPLAY-ATTACK, wherein the symbol \# denotes the number of videos.

 \begin{table}[!t]
 \begin{center}
 \caption{A summary of REPLAY-ATTACK corpus.}
 \label{tab:database}
 \resizebox{8.7cm}{!}{
 \begin{tabular}{|c|c|c|c|}
 \hline {Type}           &{Training (\#)} &{Development (\#)}  &{Test (\#)} \\
 \hline
 \hline Real-access      &60             &60                 &80       \\
 \hline Print-attack     &60             &60                 &80     \\
 \hline Phone-attack     &120            &120                &160     \\
 \hline Table-attack     &120            &120                &160     \\
 \hline  Total           &360            &360                &480     \\
 \hline
 \end{tabular}}
 \end{center}
 \end{table}

\subsection{Evaluation Metric}
% The same as speech anti-spoofing~\cite{jichenTALSP2020,jichenTIFS2020,jichenTALSP2023device} , equal error rate (EER) is selected as the evaluation metric to evaluate the performance of the proposed AAViT on FAS.  

We use equal error rate (EER) as the evaluation metric to report the results in this work. EER is defined as the point in the detection error trade-off (DET) curve, where the false alarm rate (FAR) (also known as false rejection rate) equals to miss detection rate (MDR) (also known as false acceptance rate) at a particular threshold. The FAR represents the rate between the number of spoofed faces that are judged as real faces and the total number of spoofed faces, while MDR represents the rate between the number of real faces which are judged as the spoofed face and the total number real faces. Let us consider $\alpha$ is the threshold point for EER, and then FAR ($\alpha$), MDR ($\alpha$) stand for FAR at threshold $\alpha$, MDR at threshold $\alpha$, we can then have the following:

\begin{align}
FAR(\alpha) =\frac{N_{SJR}}{ N_S}
\label{tab:far}
\end{align}
\begin{align}
MDR(\alpha) =\frac{N_{RJS}}{N_R}
\label{tab:mdr}
\end{align}
\begin{align}
FAR(\alpha)=MDR(\alpha) =EER(\alpha)
\label{tab:EER}
\end{align}
where $N_{SJR}$, $N_S$, $N_{RJS}$ and $N_R$ stand for
the number of spoofed faces that are judged as real faces, the total number of spoofed faces, the number of real faces that are judged as spoofed faces and the total number of real faces, respectively.

Again, we note that the half total error rate (HTER) used in the previous work~\cite{replayattack} is defined as:
\begin{align}
HTER(\alpha)=\frac{FAR(\alpha)+MDR(\alpha)}{2}
\label{tab:hter}
\end{align}
Thus, it can be found that $HTER(\alpha)$ equals $EER(\alpha)$ at the threshold $\alpha$ by comparing the Equations~(\ref{tab:EER}) and~(\ref{tab:hter}). 

\subsection{Experimental Setup}
The various parameters used for the transformer model in this work are the same as those considered in~\cite{vit}. All the videos are used to extract face images according to 25 frames per second, and then $256 \times 256 \times 3$ raw RGB feature is extracted for every frame face image in the training, development and test subsets.

\section{Results and Analysis}
\label{seciv}

% \subsection{Performance of proposed AAViT}
We are first interested to look at the performance of the proposed AAViT to prevent spoofing attacks on Replay-Attack database. Table~\ref{tab:exeresult} reports the the performance of AAViT on the developments as well as test set using raw RGB feature in terms of EER. It can be observed that AAViT performs very effectively as an anti-spoofing system to handle the spoofing attacks. Next, we perform some studies to compare the performance of AAViT to traditional ViT and some other known existing systems in the following subsections.

 \begin{table}[!b]
 \begin{center}
 \caption{Performance in EER (\%) of the proposed AAViT on REPLAY-ATTACK database development and test set using raw GRB feature as the input.}
 \vspace{2mm}
 \label{tab:exeresult}
 {
 \begin{tabular}{|c|c|c|c|}
 \hline\multirow{2}{*} {Feature}           &\multirow{2}{*}{Model}         &\multicolumn{2}{c|}{EER} \\
 \cline{3-4}                               &                               &Development             &Test   \\
 \cline{3-4}
 \hline
 \hline Raw RGB                    &AAViT                          &1.87                    &1.71       \\
 \hline
 \end{tabular}}
 \end{center}
 \end{table}

% As shown in Table~\ref{tab:exeresult}, the EER can reach 1.87\% and 1.7\% on the development set and test set, respectively.

\subsection{Ablation experiment: AAViT vs. ViT}

As discussed earlier, the modules of adaptive-avg-pooling and attention are the core of the proposed AAViT. Therefore, we are keen to know the significance of these two modules in the proposed AAViT. To this end,  ablation studies are performed and corresponding results on the Replay-Attack database test set are reported in Table~\ref{tab:exeablation}. Here, we compare AAViT against the traditional ViT and AAViT without (w/o) attention module. It is observed from Table~\ref{tab:exeablation} that use of adaptive-avg-pooling on ViT, i.e., AAViT w/o attention helps to improve the performance on the test by absolute 2.11\% in EER. Then on introducing the attention module further improves the results by absolute 0.48\% to obtain an EER of 1.71\% for AAViT. This signifies the impact of both adaptive-avg-pooling and attention modules in AAViT to make it effective. 

% As mentioned above, we know that the modules of adaptive-avg-pooling and attention are the core of the proposed AAViT. Here, we want to know the role of the modules of adaptive-avg-pooling and attention playing in the proposed AAViT. To this end,  ablation experiments are performed and corresponding experimental results on the test set in terms of EER are given in Table~\ref{tab:exeablation}, herein, the AAViT without (w/o) the module of attention can be named as AAViT w/o attention.

\begin{table}[!t]
\begin{center}
 \caption{Performance comparison in EER (\%) for the proposed AAViT, AAViT without (w/o) the attention module and baseline ViT using raw RGB feature as the input on REPLAY-ATTACK database test set.}
 \vspace{2mm}
 \label{tab:exeablation}
 {
 \begin{tabular}{|c|c|c|c|}
 \hline  Feature                            &Model                   &EER        \\
 \hline
 \hline \multirow{3}{*} {Raw RGB}   &AAViT                   &1.71       \\
 \cline{2-3}                                &AAViT w/o attention     &2.19         \\
  \cline{2-3}                               &ViT                     &4.30         \\
 \hline
 \end{tabular}}
 \end{center}
 \end{table}

\subsection{Comparison with commonly used models}

In this subsection, we would like to compare the proposed AAViT with some commonly used models. We consider ResNet18, ResNet50 and ResNet100 for this comparison. The results on the test set of Replay-Attack database for this comparison are reported in Table~\ref{tab:execomp}. It is noted that all the systems use raw RGB feature as the input. From Table~\ref{tab:execomp}, it can be seen that deeper the ResNet architecture, it helps to achieve an improved performance. However, AAViT outperforms different ResNet configurations including ResNet100, which has much larger model size than ResNet18 and ResNet50. This projects AAViT as a very strong anti-spoofing system for face recognition.

\begin{table}[!t]
\begin{center}
 \caption{Performance comparison in EER (\%) between the proposed AAViT and the commonly used models using raw RGB feature as the input on the REPLAY-ATTACK database test set.}
 \vspace{2mm}
 \label{tab:execomp}
 {
 \begin{tabular}{|c|c|c|c|}
 \hline  Feature                            &Model             &EER   \\
 \hline
 \hline \multirow{4}{*} {Raw RGB}   &ResNet18          &22.23       \\
 \cline{2-3}                                &ResNet50          &6.70        \\
 \cline{2-3}                                &ResNet100         &2.66        \\
 \cline{2-3}                                &AAViT             &1.71      \\
 \hline
 \end{tabular}}
 \end{center}
 \end{table}

% As seen from Table~\ref{tab:execomp}, several conclusions can be found:
% \begin{itemize}
% \item For ResNet, the deeper layer, the better performance.
% \item The proposed AAViT can perform better than the commonly used models such as ResNet50 and ResNet100.
% \end{itemize}
% Totally speaking, it can confirm that the proposed method is correct.

\subsection{Comparison with some known systems}

In this subsection, we are interested to compare the performance of the proposed AAViT with the performance of other existing systems that are reported on Replay-Attack test set. The performance other systems compared here are reported in HTER, which is equivalent to EER and Table~\ref{tab:syscomp} shows this comparison. It is noted that these systems use different feature representations as observed from Table~\ref{tab:syscomp}. In~\cite{replayattack}, local binary pattern (LBP) histogram and its variations LBP$_{3x3}$ and LBP$^{u2}$ are used, where they denote the most simple LBP pattern and uniform LBP, respectively. On the other hand, the authors of~\cite{tifs2023} consider a feature by combining embedding-level and prediction-level consistency regularization with raw RGB feature that is referred to as EPCR in short. The classifiers are also different in respective works that include linear discriminant analysis (LDA), support vector machine (SVM), central difference convolutional network (CDCN), ResNet18 and transformer. The performance comparison of the proposed AAViT based system with all other systems discussed above in Table~\ref{tab:syscomp} reveals that AAViT is much more effective than the existing systems for face anti-spoofing.

% denotes the most simple LBP pattern, LBP$^{u2}$ denotes uniform LBP, EPCR stands for feature by combining embedding-level and prediction-level consistency regularization with raw RGB featue, SSDG denotes single-side domain generalization, CDCN stands for central difference convolutional networks,

% the proposed AAViT-based system is compared with some known systems in terms of HTER. The experimental results on the test set comparison between the proposed system and some known systems in terms of HTER is given in Table~\ref{tab:syscomp}. Herein, LBP stands for local binary pattern histogram, LBP$_{3x3}$ denotes the most simple LBP pattern, LBP$^{u2}$ denotes uniform LBP, EPCR stands for feature by combining embedding-level and prediction-level consistency regularization with raw RGB featue, SSDG denotes single-side domain generalization, CDCN stands for central difference convolutional networks, 

% From Table~\ref{tab:syscomp}, it can be found that the proposed system outperforms some known systems on the test set with a little margin, we attribute that AAViT is used as the classifier in the proposed system.

\begin{table}[!t]
\begin{center}
 \caption{Performance comparison in HTER (\%) between the proposed system and some known systems on REPLAY-ATTACK database test set.}
 % \vspace{2mm}
 \label{tab:syscomp}
 {
 \begin{tabular}{|c|c|c|c|}
 \hline System                 &Feature          &Model    &HTER   \\
 \hline \hline Chingovska et al.~\cite{replayattack}&LBP$_{3x3}^{u2}$ &LDA      &17.17  \\
 \hline Chingovska et al.~\cite{replayattack}&LBP$_{3x3}^{u2}$ &SVM      &15.16  \\
 \hline Chingovska et al.~\cite{replayattack}&LBP              &SVM      &13.87   \\
 \hline Wang et al.~\cite{tifs2023}    &EPCR             &CDCN     &13.50    \\
 \hline Wang et al.~\cite{tifs2023}    &EPCR            &ResNet18  &11.38   \\
 \hline Wang et al.~\cite{tifs2023}    &EPCR            &Transformer &6.50    \\
\hline Proposed          &Raw RGB     &AAViT    &1.71   \\
 \hline
\end{tabular}}
\end{center}
\end{table}

\section{Conclusion}
\label{conc}

In this work, we proposed a novel modified transformer to deal with the issue of information loss due to the use of computing the average value in the MLP for the traditional ViT. The average value computing module is replaced by modules of adaptive-avg-pooling and attention for the modified transformer version, which is AAViT. We study the proposed modified transformer model for face anti-spoofing studies and from the results on Replay-Attack corpus the proposed AAViT emerges as a very effective system to handle spoofing attacks for face recognition. In addition, we found AAViT outperforms many commonly used state-of-the-art systems and other known systems on Replay-Attack corpus test set. The future work will focus on exploring different front-ends for AAViT and extend the model for other applications. 

% in order to escaping some information be lost in the process of computing the average value in the MLP of the traditional ViT. A modified ViT is proposed by using the modules of adaptive-avg-pooling and attention to replace the module of average value computing of ViT, we named it as AAViT. The experimental results show the proposed AAViT perform better the traditional ViT,  in equal error rate can be absolutely reduced. In addition, we also found that the proposed AAViT can perform much better than some commonly used neural networks such ResNet.

\newpage
\footnotesize
\balance
\bibliographystyle{IEEEbib}
\bibliography{refs}

\end{document}